\documentclass [amsfonts,prl,preprint]{revtex4}
\begin{document}
\title{A single-frequency test for one-parameter models of the linear
  thermo-visco-elastic response of glass-forming liquids.}

\author{ N. L. Ellegaard, T. Christensen, N. B. Olsen, and J. C. Dyre}
\affiliation{Department of Mathematics and Physics (IMFUFA), Roskilde
  University, P.O.Box 260, DK-4000 Roskilde, Denmark}
\email{gnalle@ruc.dk}

\newcommand{\gdag}{\mathfrak{G}}
\newcommand{\ndots}{...}
\newcommand{\ntub}[2]{#1_1,#1_2\ndots,#1_#2}
\newcommand{\df}[2]{{\frac{d#2}{d#1}}}
\newcommand{\pf}[2]{{\frac{\partial#2}{\partial#1}}}
\newcommand{\pftsp}[2]{\tsp{\pf{#1}{#2}}}
\newcommand{\pfr}[2]{\left .{\frac{\partial#2}{\partial#1}}\right|_0}
\newcommand{\pfrto}[2]{\left .{\frac{\partial^2#2}{\partial^2#1}}\right|_0}
\newcommand{\pfrdef}{\Big|_0}
\newcommand{\pt}[3]{\left(\frac{\partial#2}{\partial#1}\right)_{#3}}
\newcommand{\ptto}[3]{\left(\frac{\partial^2#2}{\partial#1^2}\right)_{#3}}
\newcommand{\pfto}[2]{{\frac{\partial^2#2}{\partial#1^2}}}
\newcommand{\po}[1]{\frac{\partial}{\partial#1}}
\newcommand{\pfij}[3]{\frac{\partial^2#3}{\partial#1 ~\partial#2}}
\newcommand{\pfrij}[3]{\left .\frac{\partial^2#3}
    {\partial#1 ~\partial#2}\right|_0}
\newcommand{\pref}[1]{(\ref{#1})}
\newcommand{\vektor}[1]{\left(\begin{array}{c}#1\end{array}\right)}
\newcommand{\matrice}[2]{\left(\begin{array}{#1}#2\end{array}\right)}
\newcommand{\vek}[1]{{\bf #1}}
\newcommand{\mat}[1]{{\bf #1}}
\newcommand{\dd}[1]{\delta {\tilde{#1}}}
\newcommand{\dvek}[1]{{\dd{\vek #1}}}
\newcommand{\ptc}[4] {\left(\frac{\dd #2}{\dd #1}\right)_{\dd #3\dd
    #4}}
\renewcommand{\dag}{,}

\newcommand{\lr}[1]{\left({#1}\right)}
\newcommand{\midl}[1]{{\left<#1\right>}}
\newcommand{\half}{\frac 1 2}
\newcommand{\ratio}{\Lambda}

\newcommand{\mgamma}{\Gamma^{0}}
\newcommand{\mprob}{P^{0}}
\newcommand{\improb}{\lr{P^{0}}^{-1}}
\newcommand{\mprobhalf}{\lr{P^{0}}^{\frac 1 2}}
\newcommand{\improbhalf}{\lr{P^{0}}^{-\frac 1 2}}
\newcommand{\press}{p}
\newcommand{\lll}{{l}}
\newcommand{\jj}{{j}}
\newcommand{\kk}{{k}}
\newcommand{\nn}{{n}}
\newcommand{\mm}{{m}}
\newcommand{\na}{{\alpha}}
\newcommand{\nb}{{\beta}}
\newcommand{\nc}{{\gamma}}
\newcommand{\nd}{{\delta}}
\newcommand{\dq}{\delta^Q}
\newcommand{\dv}{\delta^X}
\newcommand{\linf}{L^{\infty}}
\newcommand{\ainv}{a^{-1}}
\newcommand{\mk}[1]{#1^*}
\newcommand{\dmk}[1]{\delta {\tilde{#1}}^*}

\newcommand{\inv}{\mathfrak{I}}

\begin{abstract}
  A master equation description of the inherent dynamics is used to
  calculate the frequency-dependent linear thermo-visco-elastic
  response functions of a glass-forming liquid. From the imaginary
  parts of the isobaric specific heat, isothermal bulk modulus, and
  isobaric thermal expansion coefficient, we define a quantity
  $\Lambda_{Tp}(\omega)$ with the property that
  $\Lambda_{Tp}(\omega)=1$ is equivalent to having a one-parameter
  description of the linear thermo-visco-elastic response. This
  provides an alternative to the well-known criterion based on the
  Prigogine-Defay ratio.
\end{abstract}
\maketitle{}

[Preliminary version]

Several articles have examined if it is possible to give a one
parameter description of the thermodynamic relaxation processes of
viscous liquids
\cite{prigogine54+chemic,davies52,davies53+therm,roe77+therm,moynihan78,moynihan81+compar,nieuwenhuizen97+ehren+glass}.
The so called Prigogine-Defay ratio
\cite{prigogine54+chemic,davies52,davies53+therm,roe77+therm} tests
for a one parameter description by comparing the responses of the
liquid state with the responses of a glassy state. This approach is
problematic because the responses of the glassy state depends on the
history of the glass. In other words the original Prigogine-Defay
ratio is not an state function.  Later tests compare the equilibrium
(zero frequency) responses of a liquid to the instantaneous (infinite
frequency) responses of the same liquid
\cite{moynihan78,moynihan81+compar}.  In principle this should lead to
an test quantity which is a state function. However instantaneous
responses very difficult to measure, as one has to deal with
experimental problems which are related to heat flow and thermal
stresses in the sample.

In the following we propose a new test which allows testing for a one
parameter hypothesis by measuring the imaginary parts of three complex
response functions at a single frequency.  This way we can completely
avoid the measuring of an instantaneous response.  We examine a
harmonic linear perturbation of a thermoelastic system with
temperature $T_0$, pressure $p_0$, entropy $S_0$, and volume $V_0$.
Writing $p=p_0 + Re(\dd p~e^{i \omega t})$ and $T=T_0 +Re(\dd T~e^{i
  \omega t})$ the oscillation of $V$ and $S$ is described by $V=V_0
+Re(\dd V~e^{i \omega t})$ and $S=S_0 +Re(\dd S~e^{i \omega t})$.  The
complex coefficients $\dd p$, $\dd V$,$\dd S$, and $\dd T$ are related
by a matrix of complex, frequency dependent response functions.
\begin{eqnarray}
    \vektor{\dd S\\\dd V}
    &=&\matrice{cc}{
      \frac {V_0} {T_0}  c_p(\omega)&V_0\alpha_p(\omega)\\
       V_0\alpha_p(\omega)& V_0\kappa_T(\omega)
    }\vektor{\dd T\\ -\dd p }\label{thermoelastic}
\end{eqnarray}
Here $c_p(\omega)$ denotes the isobaric specific heat capacity per
volume, $\alpha_p(\omega)$ denotes isobaric thermal expansion
coefficient, whereas $ \kappa_T(\omega)$ denotes the isothermal bulk
compressibility. The matrix symmetric due to an Onsager relation,
which will be proven in the course of this article.

\section{Markovian inherent dynamics}
We shall base the following calculation one the so called energy
landscape picture of the dynamics of a viscous
liquid\cite{goldstein69+viscous}.  We assume that at a sufficiently
low temperature a small molecular system spends most of its time
vibrating around a minimum of the potential energy function.
Furthermore, we assume that the rate of transitions is so low that a
system achieves vibrational equilibrium between two transitions.  An
ensemble of such systems is well described by a vector of
probabilities $\vek P = (\ntub P N)$ where $P_n$ denotes the fraction
of the ensemble which is vibrating around the energy minimum $n$.  For
a large system the rate of transitions is proportional to the system
size, and therefore we cannot expect a large system to achieve
vibrational equilibrium between each transition. However assuming that
a transition from one potential minimum to another minimum is a
localized event, leads to an assertion of local vibrational
equilibrium.

In the following we examine a linear experiment in which we control
the intensive variables $T$ and $p$ and calculate the resulting change
of the extensive variables $S$ and $V$.  Here $S$ denotes the ensemble
entropy whereas $V$ denotes the ensemble average of the volume.
\begin{eqnarray}
   S &=& -\sum_{m} P_m \lr{\pf T {G_m} + k_B  \ln P_m}\\
   V &=& \sum_{m} P_m \pf p {G_m} 
\end{eqnarray}
$G_n(T,p)$ denotes the vibrational Gibbs energy of a single minimum,
whereas the Gibbs energy of an ensemble is given by
\begin{eqnarray}
  G(T,p,\vek P)  &=&  \sum_n P_n 
  \lr{G_n(T,p)  + k_B T \ln P_n}. \label{gibbs}
\end{eqnarray}
We introduce the notation $\vek X=(T,\vek -p)$ and $\vek Q= (S,V)$,
and note that the following equation is fulfilled for all choices of
$\vek P$
\begin{eqnarray}
   Q_\na &=& -\pf{X_\na}{G},~~~\na = 1,2.\label{defq}
\end{eqnarray}
The equilibrium distribution is easily found by minimizing $G$ under
the constraint $\sum_m P_m=1$%
\footnote{The relation $\pf {P_n} G = const$ leads to $G_n + k_B T\ln
  P_n^{eq} + k_BT = const$. Insert in \pref{gibbs} to find $G_n + k_B
  T\ln P_n^{eq} = G$, and isolate $P_n^{eq}$ to get the desired
  result}.
\begin{eqnarray}
   P_n^{eq}(T,p)&=&\exp\lr{-\frac{G_n - G}{k_BT}}. \label{peq}   
\end{eqnarray}
For a linearly perturbed system, the rate of entropy change, $\dot S$,
is related to the heat flow, $J_q$, by the relation $J_q ~=~ T\dot S$
\footnote{We define the enthalpy from the Gibbs energy . $ H(T,p,\vek
  P) ~=~ G - T \pf T G$. Inserting  \pref{gibbs} we find that $H$
  corresponds to the average vibrational entropy $ H(T,p,\vek P) ~=~
  \sum_n P_n \lr{G_n - T \pf T {G_n}} $. We define $\delta q = \delta
  H - V \delta p =\delta H - \pf p G \delta p$. A little calculation
  yields $\delta q ~=~ -T \pfto T G \delta T - T \pfij T p G \delta p
  - T \pfij T {P_n} G \delta P_n ~=~T \delta S $ }.
Thus the rate of change of the energy is given by $\dot U = T\dot S
-p\dot V$. We insert in the relation $G = U -TS+pV$ to get
\begin{eqnarray}
\dot G &=& -S \dot T + V \dot p~=~-\sum_\na Q_\na\dot X_\na.\label{dotg}
\end{eqnarray}
When the thermal energy per degree of freedom is low compared the
height of an energy barrier, the transitions between minima is well
described as a Markov process, and the dynamics of $P_{m}$ it is
described by a master equation \cite{kampen81+stoch}.
\begin{eqnarray*}
  \dot P_n &=&\sum_m W_{n m} P_{ m}  \label{master}
\end{eqnarray*}
The rate matrix $W$ depends on $T$ and $p$, and the equilibrium
distribution $P^{eq}_m$ fulfills $\sum_m W_{n m}P^{eq}_ m=0$.  We
proceed to examine linear perturbations around a reference state $\vek
X^0 = (T^0,-p^0)$.  Furthermore we introduce $W_{ m n}^0=W_{ m
  n}(T_0,p_0)$ and $P_{ m}^{0} =P_{ m}^{eq}(T_0,p_0)$. We let $\delta$
denote perturbations around the reference state to get
\begin{eqnarray*}
  \delta \dot P_n &=&\sum_m W_{n m}^0 \delta P_{ m} + \delta W_{n m} P_{ m}^{0}.
\end{eqnarray*}
We note that $\delta( \sum_m W_{n m} P_{ m}^{eq})=0$. This leads to
$\sum_m \delta W_{n m} P_{ m}^0 =-\sum_m W_{n m}^0 \delta P_{
  m}^{eq}$. We insert to get
\begin{eqnarray}
  \delta \dot P_n &=&\sum_m W_{n m}^0 \lr {\delta P_{ m}-\delta P_{ m}^{eq}}.
  \label{wdp}
\end{eqnarray}
%
%We would like to expand $\delta P_{ m}^{eq}$ in terms of $\delta
%X_\na$.  
A little algebra and the use of \pref{peq} leads to
\footnote{First note $\pf {(-p)}{\ln P_ m^{eq}}=\frac {-1} {k_BT_0}\pf
  {(-p)}{G_ m -G}=\frac {1} {k_BT_0}\lr{\pf{P_m}V - V}$. Then note
  $\pf {T}{\ln P_ m^{eq}} = 
  \frac {-1} {k_B T} \lr{\pf {T}{G_ m -G} - \frac{G_m -G} {T}}
  = \frac {-1} {k_B T}\lr{\pf T{G_m} -\pf T{G} + k_B \ln P_m^{eq}}
%  = \frac {-1} {k_B T}\lr{\pf T{G_m +k_B T\ln P_m^{eq} -G}}
%  = \frac {-1} {k_B T}\lr{\pf T{G_m +k_B T\ln P_m^{eq} -G}}
%  =\frac {-1} {k_B T}\lr{\pfij {P_m} T G - \pf T G - k_B}
  =\frac 1 {k_B T}\lr{\pf {P_m} S - S + k_B}$ 
}
\begin{eqnarray}
\pf{X_\nb}{\ln
  P_{ m}^{eq}}&=&
\frac {1}{k_BT_0 }\lr{
\pf {P_m} {Q_{\nb}} -Q_{\nb} + \delta_{\nb 1} k_B}.
\end{eqnarray}
We insert in \pref{wdp} and remember $\sum_m W_{n m}^{0} P^0_{ m}=0$
to get
\begin{eqnarray}
\delta \dot  P_n 
&=& \sum_m W_{n m}^{0} \lr{\delta  P_ m- \sum_\nb\frac 1{k_BT_0} P^0_{ m}\pf {P_m} {Q_{\nb}}\delta X_\nb
}.\label{dotp}
\end{eqnarray}
The steady state fulfills
\begin{eqnarray}\sum_n 
\lr{ W_{ln}^0-i\omega\delta_{ln}} \dd P_n 
&=& \sum_{m,\nb}
 \frac {1}{k_BT_0}W_{l m}^0  P_ m^0\pf {P_m} {Q_{\nb}}\dd X_\nb.
\label{laplace}
\end{eqnarray}
Here $ \dd P_n$ and $\dd X_\nb$ denote complex amplitudes.  For
$\omega>0$ we can simplify \pref{laplace} by defining
\begin{eqnarray}
  A_{n m}(\omega) &=& \frac {1} {k_BT_0}
  \sum_l\lr{ W^0 -i\omega I }^{-1}_{nl}
  W_{l m}^0 P_m^0 .\label{akrav}
\end{eqnarray}
Here $I$ denotes the identity matrix. We substitute \pref{akrav} into
\pref{laplace} to get
\begin{eqnarray}
\dd P_n  
&=& \sum_{m,\nb} A_{n m}(\omega) \pf {P_m} {Q_{\nb}} \dd X_\nb.\label{a}
\end{eqnarray}
This equation describes the dynamics of the ensemble. In order to
calculate the response functions, we define $J_{\na\nb}^{\infty}=
-\pfij {X_\na}{X_\nb} G$ and expand the complex amplitudes $\dd Q_\na$
to the first order to get
\begin{eqnarray}
  \dd Q_\na 
&=& \sum_{n}\pf {P_n} {Q_{\na}} \dd P_n 
+ \sum_{\nb}J_{\na\nb}^{\infty} \dd X_\nb.\label{ddq}
\end{eqnarray}
We insert in \pref{a} to get an expression for the thermoviscoelastic
response functions that were introduced in \pref{thermoelastic}. This
gives $\dd Q_\na = \sum_{\nb}J_{\na\nb}(\omega)\dd X_\nb$, where
\begin{eqnarray}
  J_{\na\nb}(\omega)
  &=&J_{\na\nb}^{\infty} 
  +\sum_{m,n}\pf {P_n} {Q_{\na}}  A_{n m}(\omega)  \pf {P_m} {Q_{\nb}}.
  \label{dq}
\end{eqnarray}
As $A_{n m}(\infty) =0$ we get $J_{\na\nb}(\infty)=J_{\na\nb}^\infty$.
We can show that $J_{\na\nb}(\omega)$ is symmetric by introducing a
symmetric matrix $Y_{mn}$ with the elements $Y_{mn} =
(P_{m}^0)^{-\half} W_{mn} (P_{n}^0)^\half$ \footnote{The symmetry of
  $Y$ follows from detailed balance $W_{mn} p^{eq}_n =W_{nm}
  p^{eq}_m$}. This gives
\begin{eqnarray}
  A_{nm}(\omega) &=& \frac {1} {k_BT_0}\sum_{l}
  (P_{n}^0)^{\half}\lr{i\omega I + Y}^{-1}_{nl} Y_{lm}
  (P_{m}^0)^{\half}.\label{gamma}
\end{eqnarray}
The matrices $\lr{i\omega I + Y}^{-1}$ and $Y$ commute so it is easy
to show that $A_{nm}(\omega)$ is symmetric. We insert in \pref{dq} to
conclude that the matrix $J(\omega)$ is also symmetric.

\section{Prigogine-Defay ratio}
Having calculated the response functions we can insert in the
``linear Prigogine-Defay ratio''
\cite{roe77+therm,moynihan78,moynihan81+compar}
\begin{eqnarray}
  \Pi &=& \frac{
    ( c_p(0)-  c_p(\infty))
    ( \kappa_T(0)-  \kappa_T(\infty))}   
  {T_0(\alpha_p(0) -\alpha_p(\infty))^2}. \label{prigogine}
\end{eqnarray}
We insert \pref{dq} and remember $A_{n m}(\infty)=0$ to get
\begin{eqnarray}
   \Pi 
&=& \frac{
\lr{\sum_{m,n}\pf{P_n}{S}A_{n m}(0)\pf{P_m}{S}}
\lr{\sum_{m,n}\pf{P_{n}}{V}A_{n m}(0) \pf{P_{m}}{V}}
}
{ 
\lr{\sum_{m,n}\pf{P_{n}}{S}A_{n m}(0) \pf{P_{m}}{V}}^2
} \label{prigogineto}.
\end{eqnarray}
From the Schwartz inequality to find $\Pi\geq 1$. We can describe the
special case $\Pi=1$ by noting that null space of $A_{ mn}(\omega)$ is
given by $\mathrm{null}(A(\omega)) =\mathrm{span} \{e\}$. Here $e$ is
a vector with elements on the form $e_n = \frac 1 N$ \footnote{ From
  the definition of equilibrium we get $\sum_{n} W_{mn}^0P_n^0=0$.
  This gives $\sum_n A_{mn}(\omega)=0$. The remaining eigenvalues are
  all positive and real \cite{kampen81+stoch}. We can find the
  spectrum by performing an orthogonal diagonalization of $Y$ to get
  $Y_{lm} =\sum_k S_{l k}\lambda_k S_{mk}$ We insert in \pref{gamma}
  to find $A_{ mn}(\omega) = \frac 1 {k_BT_0}\sum_{k}(P_{m}^0)^{\half}
  S_{m k}\frac{ \lambda_k}{ \lambda_k-i \omega} S_{n k}
  (P_{n}^0)^{\half}$}.  We note that $\Pi=1$ is only fulfilled we can
define constants $\gamma$ and c, so that all the following equation is
fulfilled for all $n$
\begin{eqnarray}
\pf{P_n}{V}&=&\gamma \pf{P_n}{S} + c.\label{parallel}
\end{eqnarray}
This result is similar to earlier treatments of the Prigogine-Defay
ratio.  As mentioned in the introduction, the Prigogine-Defay ratio is
very difficult to measure, and therefore we propose the following
ratio which is inspired by \cite{meixner59+therm,lesikar80+some}
\begin{eqnarray*}
  \ratio_{Tp}(\omega)&=&\frac{
    \frac 1 {T_0} c_p''(\omega)\kappa_T''(\omega)}
  {(\alpha_p''(\omega))^{2}}.
\end{eqnarray*}
We insert \pref{dq} to get
\begin{eqnarray*}
\ratio_{Tp}(\omega)
&=& \frac{
\lr{\sum_{m,n}\pf{P_n}{S}A_{n m}''(\omega)\pf{P_m}{S}}
\lr{\sum_{m,n}\pf{P_{n}}{V}A_{n m}''(\omega) \pf{P_{m}}{V}}
}
{ 
\lr{\sum_{m,n}\pf{P_{n}}{S}A_{n m}''(\omega) \pf{P_{m}}{V}}^2
} 
\end{eqnarray*}
From the Schwartz inequality we find that $\ratio_{Tp}(\omega)\geq 1$,
for all $\omega \in ]0,\infty[$. In order to analyze the reverse
relation we note that $\mathrm{null}(A''(\omega)) = \mathrm{span}\{e\}$.  This
tells us for all $\omega \in ]0,\infty[$ that $\ratio_{Tp}(\omega)= 1$
is equivalent to \pref{parallel} and thus $\ratio_{Tp}(\omega)= 1$ is
equivalent to $\Pi=1$. We can get a better understanding of the
implications of \pref{parallel} if we approximate $G_n$ by a Taylor
expansion in $T$ to get $G_n = H_n - T S^{vib}_n$. This
approach is essentially equivalent to a series of recent articles on
thermodynamics of the energy landscape
\cite{nave02+poten,nave03+numer,shell03+energ}. A little calculation
yields \footnote{From \pref{defq} we get $\pf{P_n}{S}=-\pfij{P_n}T G = - \pf
  T {G_n} - k_B\lr{ 1 + \ln P_n}$. Applying this formula to the
  equilibrium ensemble \pref{peq}, we get $\pf{P_n}{S}=-\pf T {G_n}
  +\frac{G_n - G -k_B T} T=\frac{H_n - G -k_B T} T$}
\begin{eqnarray}
 \pf{P_n}{S}  &=& \frac{H_n - G -k_B T_0} {T_0}. \label{un}
\end{eqnarray}
We insert in \pref{parallel} to get
\begin{eqnarray}
\pf {P_n} V 
&=& \gamma \frac {H_n} {T_0}  + \lr{c - \gamma ~\frac{G + k_BT_0}{T_0}} \label{eis}
\end{eqnarray}
This formula shows that if $\Pi=1$ and if $G_n = H_n - T S^{vib}_n$,
then the equilibrium fluctuations of inherent enthalpy are directly
correlated to the fluctuations of inherent volume.
\appendix
\section{Single parameter model}
In the following we will describe a simple formalism that makes it
easy to perform calculations on a model fulfilling $\Pi=1$.  This
calculation is inspired by the circuit model formalism for
thermodynamics 
\cite{oster73+networ,peusner86+studies,peusner90+noneq,mikulecky93+applic}.
Inspired by Goldstein and La Nave et. al. we introduce the following
parameter \cite{goldstein63+some,nave02+poten}
\begin{eqnarray}
  \delta \varepsilon &=&\pf{P_n}{S}\delta P_n~=~
  \gamma^{-1}\pf{P_m}{V}\delta P_m.
    \label{dimarzio}
\end{eqnarray}
We combine \pref{dimarzio} with \pref{ddq} to find that in a linear
experiment the instantaneous values of $S$ and $V$ are functions of
$T$,$p$ and $\delta\varepsilon$.
\begin{eqnarray}
 \delta  S &=&  \delta\varepsilon+
J_{11}^\infty \delta T
 - J_{21}^\infty  \delta p\nonumber\\
 \delta V &=&  \gamma \delta\varepsilon+
   J_{12}^\infty \delta T
 -J_{22}^\infty \delta p
 \label{order}
\end{eqnarray}
This model is different from the order parameter models introduced by
Meixner and Prigogine, as it allows for $\delta\varepsilon$ to have non
exponential relaxation \cite{meixner59+therm,prigogine54+chemic}.  In
the following we will show that the dynamics of this system can be
described as the flow of Gibbs energy between the two reservoirs 1 and
2 with the Gibbs energy densities ${G_1}$ and $G_2$.  These reservoirs
will be chosen so that the total Gibbs energy $G$ is given by
\begin{eqnarray}
  G &=& {G_1} + G_2\label{gplusg}
\end{eqnarray}
We define the function ${G_1}$ to be the loss Gibbs energy during an
isostructural process leading the system, from a state $(T,p,\bar P)$
to the state $(T_0,p_0,\bar P)$.
\begin{eqnarray}
  {G_1}
  &=& \int_{T_0,p_0,\bar P}^{T_0,p,\bar P}
  V~ dp^*  
  - \int_{T_0,p,\bar P}^{T,p,\bar P}
   S~ dT^*\label{gdag}
\end{eqnarray}
We insert \pref{order} and $J_{12}^\infty=J_{21}^\infty$ to find $G_1$
is a function of $\delta T,\delta p$ and $\delta\varepsilon$.  Furthermore
we get the following second order expansion.
\begin{eqnarray*}
  {G_1}(\delta T,\delta p,\delta\varepsilon)  
  &=&V_0 \delta p - S_0 \delta T\\
  &&+\frac 1 2
    \vektor{\delta T\\-\delta p \\\delta \varepsilon}^T
     \matrice{ccc}{
J_{11}^\infty&J_{12}^\infty&1\\
J_{21}^\infty&J_{22}^\infty&\gamma\\
%     c_p(\infty)&\alpha_p(\infty)&\gamma\\
%     \alpha_p(\infty)&\kappa_T(\infty)&1\\
%     \pfr T \sdag&\pfr T \vdag&\gamma\\
%     \pfr T \vdag&-\pfr p \vdag&1\\
     1&\gamma&0}
    \vektor{\delta T\\-\delta p \\\delta\varepsilon} 
\end{eqnarray*}
We would like to derive an equation of conservation of Gibbs energy
 for reservoir 1. Therefore we define $\psi = - \pf
\varepsilon {G_1}$ and note from \pref{gdag} that $S = - \pf T {G_1}$
and $V = \pf P {G_1}$ to get
\begin{eqnarray}
 \dot {G_1}  &=& -S \dot T + V \dot p - \psi \dot \varepsilon.
\label{dotgdag}
\end{eqnarray}
A similar equation for reservoir 2 follows from
\pref{dotg},\pref{gplusg} and \pref{dotgdag}
\begin{eqnarray}
 \dot G_2  &=&  \dot G - \dot G_1 ~=~
\psi \dot \varepsilon \label{dotgto}
\end{eqnarray}
Most of the literature on network thermodynamics is expressed in terms
of energy flows, and in order to stay consistent with this literature
we define reservoir energies $U_1 = {G_1} + ST - pV$, and $U_2 = G_2$
to get
\begin{eqnarray}
  U &=& U_1 + U_2\nonumber\\
 \dot U_1  &=& T \dot S - p  \dot V - \psi \dot \varepsilon
 \label{dotu}\\
 \dot U_2  &=&  \psi \dot \varepsilon\nonumber
\end{eqnarray}
These formulas describe the dynamics of the liquid as energy flow
between two reservoirs. We proceed to find equations for the dynamics
of each reservoir.  The dynamics of the first reservoir follows from
\pref{gdag}. We use the definitions of $S$, $V$ and $\psi$ and examine
a harmonic perturbation to get
\begin{eqnarray}
 \vektor{\dd S \\ \dd V \\-\dd \psi} 
  &=&
     \matrice{ccc}{
J_{11}^\infty&J_{12}^\infty&1\\
J_{21}^\infty&J_{22}^\infty&\gamma\\
     -1&-\gamma&0}
    \vektor{\dd T\\-\dd p \\\dd\varepsilon}. \label{responsmatrix}
\end{eqnarray}
The partial antisymmetry of this matrix reflects the fact that $-p$,
$T$ and $-\psi$ are generalized forces on reservoir 2, whereas $V$,
$S$ and $\varepsilon$ are generalized charges. In order to describe
the second reservoir, we need a relation between $\dd \psi$ and $\dd
\varepsilon$. We use \pref{laplace}, \pref{akrav} and \pref{a} to get
$\dd \varepsilon = a(\omega) \dd \psi$, where \footnote{ From
  \pref{responsmatrix} we get $\dd \psi = \dd T - \gamma \dd p$.  Use
  \pref{parallel} to get $\pf{P_n}{S}\dd \psi ~=~ \pf {P_n} {S}\dd T -
  \gamma \pf {P_n} {S} \dd p ~=~\pf {P_n} {S}\dd T - \pf {P_n} {V} \dd
  p - c \dd p ~=~ \sum_{\na}\pf {P_n} {Q_{\na}} \dd X_{\na}- c \dd p$.
  We insert in \pref{a} and remember $\sum_m A_{n m}(\omega)=0$ to get
  $\dd P_n = \sum_m A_{n m}(\omega) \pf{P_m}{S} \dd \psi$. Finally
  insert \pref{dimarzio} to get $\dd \varepsilon = \sum_{m,n}
  \pf{P_n}{S}A_{n m}(\omega) \pf{P_m}{S} \dd \psi$ }
\begin{eqnarray}
a(\omega)\ &=& \sum_{m,n} \pf{P_n}{S} A_{n m}(\omega)  \pf{P_m}{S}. \label{aomega}
\end{eqnarray}
Finally we insert $\dd \varepsilon = a(\omega) \dd \psi$ in \pref
{responsmatrix} to find a simple expressions for the three complex
response functions that we defined in \pref{thermoelastic}.
\begin{eqnarray*}
    \frac {V_0} {T_0}  c_p(\omega)&=&
    J_{11}^\infty
    + a(\omega)\\
    V_0 \alpha_p(\omega)&=&
    J_{12}^\infty
    + \gamma  a(\omega)\\
    V_0 \kappa_T(\omega)&=& 
    J_{22}^\infty+ \gamma^2 a(\omega)
\end{eqnarray*}

%\bibliographystyle{apsrev}
%\bibliography{phd}

\end{document}